\documentclass[aps,prd,reprint,superscriptaddress,showkeys,showpacs]{revtex4-1}
\usepackage{graphicx} %
\usepackage{amsmath} %
\usepackage{amssymb} %
\usepackage{wasysym} %
\usepackage{booktabs} % \toprule \midrule \bottomrule \cmidrule(lcr){1-3}
\usepackage{lineno} %
\usepackage{hyperref} %
\hypersetup{colorlinks=true, urlcolor=blue, linkcolor=blue, citecolor=red} %
\numberwithin{equation}{section} % change equation numbering format
\usepackage{color} %
\usepackage[usenames,dvipsnames]{xcolor} %
\definecolor{gray}{gray}{0.30} %
\usepackage{float} %
\usepackage{upgreek} %
\usepackage{multirow} %

\begin{document}

%%% Some user definitions for convenience

\newcommand{\dm}{DM-Ice17}

\newcommand{\iso}[2]{\ensuremath{^{#2}\mathrm{#1}}}

\newcommand{\startdate}{July 2011}
\newcommand{\stopdate}{June 2013}

\newcommand{\dma}{Det-1}
\newcommand{\dmb}{Det-2}

\newcommand{\black}[1]{{\color{black}{#1}}}
\newcommand{\red}[1]{{\color{Red}{\textbf{#1}}}} % means doulbe check
\newcommand{\blue}[1]{{\color{Blue}{\textbf{#1}}}}
\newcommand{\green}[1]{{\color{Green}{\textbf{#1}}}} % means try to re-word
\newcommand{\gray}[1]{{\color{gray}{#1}}}
\newcommand{\orange}[1]{{\color{BurntOrange}{#1}}}

\newcommand{\h}{!ht}
\newcommand{\fw}{\textwidth}
\newcommand{\tqw}{0.75\textwidth}
\newcommand{\hw}{0.48\textwidth} %%% for 2-column view
\newcommand{\qw}{0.23\textwidth}

%%% Referencing
\newcommand{\fig}[1]{Fig.\,\ref{#1}}
\newcommand{\tab}[1]{Tab.\,\ref{#1}}
\newcommand{\sect}[1]{Sec.\,\ref{#1}}
\newcommand{\chap}[1]{Chap.\,\ref{#1}}
\newcommand{\equ}[1]{Eq.\,\ref{#1}}
\newcommand{\cit}[1]{Ref.\,\cite{#1}}

%%% UNITS
%\renewcommand{\deg}{\,C}
\renewcommand{\deg}{\ensuremath{{^{\circ}}\mathrm{C}}}

\newcommand{\aprox}{\ensuremath{\sim}\,}
\newcommand{\per}{\,\%}
\newcommand{\kpa}{\,kPa}
\newcommand{\hz}{\,Hz}
\newcommand{\s}{\,sec}
\newcommand{\lt}{\ensuremath{\,<\,}}
\newcommand{\gt}{\ensuremath{\,>\,}}
\newcommand{\plm}{\ensuremath{\,\pm}\,}
\newcommand{\kg}{\,kg}
\newcommand{\kgyr}{\,kg\ensuremath{\cdot}yr}
\newcommand{\gs}{\ensuremath{\,g}}
\newcommand{\diam}{\ensuremath{\varnothing\,}}
\newcommand{\usec}{\,\ensuremath{\upmu s}}
\newcommand{\ubk}{\,\ensuremath{\upmu}Bq/kg}
\newcommand{\mbk}{\,mBq/kg}

\newcommand{\kev}{\,\ensuremath{\mathrm{keV}}}
\newcommand{\mev}{\,\ensuremath{\mathrm{MeV}}}
\newcommand{\kevee}{\,\ensuremath{\mathrm{keV}_{\mathrm{ee}}}}
\newcommand{\mevee}{\,\ensuremath{\mathrm{MeV}_{\mathrm{ee}}}}
\newcommand{\kevr}{\,\ensuremath{\mathrm{keV}_{\mathrm{r}}}}
\newcommand{\dru}{\,\ensuremath{\mathrm{counts/day/keV/kg}}}

\newcommand{\gev}{\ensuremath{{\rm GeV}/{\rm c}^2}}
\newcommand{\smod}{\ensuremath{{S_m}}}
\newcommand{\snaught}{\ensuremath{{S_0}}}
\newcommand{\rzero}{\ensuremath{{R_0}}}
\newcommand{\nzero}{\ensuremath{{N_0}}}

\newcommand{\e}[1]{\ensuremath{\times 10^{#1}}}

\title{Measurement of Muon Annual Modulation and Muon-Induced Phosphorescence in NaI(Tl) Crystals with DM-Ice17}

% !TEX root = Muon_PRD.tex
%%%%%%%%%%%%%%%%%%%%%%%%%%%%%%%%%%%%%%%%%%%%%%%%%%%%%%%%%%%%
%%  Authors and affiliations
%%%%%%%%%%%%%%%%%%%%%%%%%%%%%%%%%%%%%%%%%%%%%%%%%%%%%%%%%%%%
%%  \author{}
%%  \affiliation{}
%%  \email{}
%%%%%%%%%%%%%%%%%%%%%%%%%%%%%%%%%%%%%%%%%%%%%%%%%%%%%%%%%%%%
\author {J.~Cherwinka}
\affiliation {Physical Sciences Laboratory, University of Wisconsin-Madison, Stoughton, WI 53589, USA}
\author {D.~Grant}
\affiliation {Department of Physics, University of Alberta, Edmonton, Alberta, Canada}
\author {F.~Halzen}
\affiliation {Department of Physics and Wisconsin IceCube Particle Astrophysics Center, University of Wisconsin-Madison, Madison, WI 53706, USA}
\author {K.~M.~Heeger}
\affiliation {Department of Physics, Yale University, New Haven, CT 06520, USA}
\author {L.~Hsu}
\affiliation {Fermi National Accelerator Laboratory, Batavia, IL 60510, USA}
\author {A.~J.~F.~Hubbard}
\email[Corresponding author: ]{antonia.hubbard@northwestern.edu}
\altaffiliation [Current affiliation: ]{Department of Physics and Astronomy, Northwestern University, Evanston, IL 60208, USA}
\affiliation {Department of Physics and Wisconsin IceCube Particle Astrophysics Center, University of Wisconsin-Madison, Madison, WI 53706, USA}
\affiliation {Department of Physics, Yale University, New Haven, CT 06520, USA}
\author {A.~Karle}
\affiliation {Department of Physics and Wisconsin IceCube Particle Astrophysics Center, University of Wisconsin-Madison, Madison, WI 53706, USA}
\author {M.~Kauer}
\affiliation {Department of Physics and Wisconsin IceCube Particle Astrophysics Center, University of Wisconsin-Madison, Madison, WI 53706, USA}
\affiliation {Department of Physics, Yale University, New Haven, CT 06520, USA}
\author {V.~A.~Kudryavtsev}
\affiliation {Department of Physics and Astronomy, University of Sheffield, Sheffield, UK}
\author {K.~E.~Lim}
\affiliation {Department of Physics, Yale University, New Haven, CT 06520, USA}
\author {C.~Macdonald}
\affiliation {Department of Physics and Astronomy, University of Sheffield, Sheffield, UK}
\author {R.~H.~Maruyama}
\email[Corresponding author: ]{reina.maruyama@yale.edu}
\affiliation {Department of Physics, Yale University, New Haven, CT 06520, USA}
\author {S.~M.~Paling}
\affiliation{STFC Boulby Underground Science Facility, Boulby Mine, Cleveland, UK}
\author {W.~Pettus}
\affiliation {Department of Physics and Wisconsin IceCube Particle Astrophysics Center, University of Wisconsin-Madison, Madison, WI 53706, USA}
\affiliation {Department of Physics, Yale University, New Haven, CT 06520, USA}
\author {Z.~P.~Pierpoint}
\affiliation {Department of Physics and Wisconsin IceCube Particle Astrophysics Center, University of Wisconsin-Madison, Madison, WI 53706, USA}
\affiliation {Department of Physics, Yale University, New Haven, CT 06520, USA}
\author {B.~N.~Reilly}
\altaffiliation {Current affiliation: Department of Physics and Astronomy, University of Wisconsin-Fox Valley, Menasha, WI 54952, USA}
\affiliation {Department of Physics and Wisconsin IceCube Particle Astrophysics Center, University of Wisconsin-Madison, Madison, WI 53706, USA}
\affiliation {Department of Physics, Yale University, New Haven, CT 06520, USA}
\author {M.~Robinson}
\affiliation {Department of Physics and Astronomy, University of Sheffield, Sheffield, UK}
\author {P.~Sandstrom}
\affiliation {Department of Physics and Wisconsin IceCube Particle Astrophysics Center, University of Wisconsin-Madison, Madison, WI 53706, USA}
\author {N.~J.~C.~Spooner}
\author {S.~Telfer}
\affiliation {Department of Physics and Astronomy, University of Sheffield, Sheffield, UK}
\author {L.~Yang}
\affiliation {Department of Physics, University of Illinois at Urbana-Champaign, Urbana, IL 61801, USA}
\collaboration{The DM-Ice Collaboration}
\date{\today}
\begin{abstract}

We report the measurement of muons and muon-induced phosphorescence in DM-Ice17, a NaI(Tl) direct detection dark matter experiment at the South Pole. Muon interactions in the crystal are identified by their observed pulse shape and large energy depositions. The measured muon rate in DM-Ice17 is 2.93\,$\pm$\,0.04\,$\upmu$/crystal/day with a modulation amplitude of 12.3\,$\pm$\,1.7\%, consistent with expectation. Following muon interactions, we observe long-lived phosphorescence in the NaI(Tl) crystals with a decay time of 5.5\,$\pm$\,0.5\,s. The prompt energy deposited by a muon is correlated to the amount of delayed phosphorescence, the brightest of which consist of tens of millions of photons. These photons are distributed over tens of seconds with a rate and arrival timing that do not mimic a scintillation signal above 2\kevee. While the properties of phosphorescence vary among individual crystals, the annually-modulating signal observed by DAMA cannot be accounted for by phosphorescence with the characteristics observed in DM-Ice17. 
\end{abstract}

\maketitle

% !TEX root = Muon_PRD.tex

\section{Introduction}
\label{sec:intro}
Astronomical evidence converges on a 27\% cold dark matter component of the universe~\cite{Planck}, but the properties of dark matter remain largely unknown. The Weakly Interacting Massive Particle (WIMP) is a theoretically motivated dark matter candidate~\cite{WIMP, Bertone}, and a number of direct detection dark matter experiments are underway~\cite{Direct_Detection} to search for evidence of WIMP-nucleon scattering~\cite{Scattering1, Scattering2}. 

The WIMP interaction rate in terrestrial-based detectors is expected to have an annual modulation due to the Earth's motion relative to the galactic WIMP halo~\cite{Modulation, Colloquium}. The only claim of dark matter detection comes from the DAMA/NaI and DAMA/LIBRA experiments at the Laboratori Nazionali del Gran Sasso (LNGS) in Italy. Combined, these NaI(Tl) detectors have observed a 9.3$\sigma$ signal over 14 annual cycles~\cite{DAMAFinal}, while other experiments have reported null results in conflict with the observed modulation~\cite{LUX, EDELWEISS, XENON, CDMS, CRESST, Panda, PICO, PICASSO, KIMS, COUPP, XENON_SD, SIMPLE}.  Several experiments aim to directly test DAMA's dark matter claim using the same target material of NaI(Tl)~\cite{DMIce, ANAISDet, KIMSNaI, SABREDet, KamPICO}.

A number of hypotheses for the origin of DAMA's annually modulating signal have been proposed, ranging from alternative dark matter models~\cite{Schnee} to potential sources of background~\cite{Ralston, Blum, Davis, Nygren}. No background has yet been shown to completely account for the DAMA modulation signal~\cite{No_role, DAMAProceedings, No_role2, FM, Klinger, Sheffield}.  Atmospheric muons are one such background, the production of which in the Earth's atmosphere is correlated with the air density in the upper atmosphere and therefore exhibits a seasonal modulation~\cite{Barrett, Ambrosio, Adamson, Agafonova, Paolo, Borexino}.  In this paper we report the measurement of the muon rate and characterization of muon-induced phosphorescence in DM-Ice17, a NaI(Tl)-based direct detection dark matter experiment currently operating at the South Pole~\cite{First_data}.
% !TEX root = Muon_PRD.tex

\section{DM-Ice17 Detector}
\label{sec:dmice}

\begin{figure}[!htb]
	\includegraphics[width=\hw]{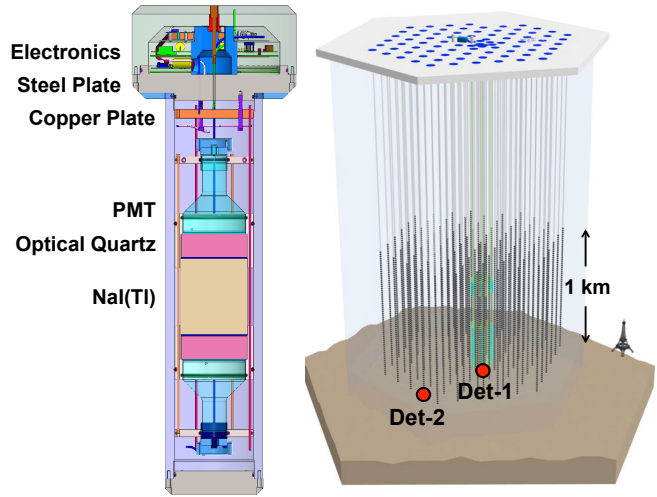}
	\caption{DM-Ice17 detector schematic (left) and the location of the DM-Ice17 detectors within IceCube (right). A stainless steel pressure vessel encloses the NaI(Tl) crystal coupled to two PMTs, two IceCube mainboards, and two high voltage boards. The DM-Ice17 detectors are deployed 2457\,m deep in the South Pole ice and 7.5\,m below the bottom IceCube modules. Det-1 is deployed below the DeepCore infill array (green volume), while Det-2 is located on the edge of the 1\,km$^3$ IceCube volume.}
	\label{DM-Ice17}
\end{figure}

%%%%%%%%%%%%%%%%%%%%%%%%%%%%%%%%%%%%%%%
DM-Ice17 consists of two identically-designed detectors, Det-1 and Det-2, deployed 2457\,m (2200\,m.w.e.) deep in the South Pole ice~\cite{First_data}. The NaI(Tl) crystals are located 7.5\,m below the bottom detectors of the IceCube Neutrino Observatory (IceCube)~\cite{IceCube, IceCube:PRL2015, IceCube:PRL2013, IceCube:PRL2013DM, IceCube:Science2013}. Det-1 is located near the center of the IceCube array while Det-2 is located on the array edge, as shown in Figure~\ref{DM-Ice17}. The glacial ice provides a high radiopurity environment, stable running conditions, and efficient neutron moderation.

DM-Ice17 has achieved an uptime of $>$\,99\% since the start of physics running in June 2011. This study uses 3.5\,yrs of data from July 1, 2011 to December 31, 2014. Day 1 is set to January 1, 2011; the data shown is from days 182 to 1460.

Each DM-Ice17 detector contains a cylindrical 8.47\,kg NaI(Tl) crystal (14.0\,cm diameter, 15.0\,cm height) coupled on each end to a photomultiplier tube (PMT). Details of the crystals' history and contamination are provided in \cite{First_data}. The high voltage and digitizing electronics are located in an isolated volume above the upper PMT. The entire setup is enclosed in a stainless steel pressure vessel.

Each DM-Ice17 PMT is independently controlled and monitored by its own IceCube mainboard (DOM-MB) assembly and high voltage (HV) board~\cite{IC_DAQ}.  Waveforms are recorded and digitized when paired PMTs trigger within 350\,ns of each other, referred to as local coincidence (``LC"). Timing is set by a clock on the surface and synchronized to the GPS receiver at the IceCube Lab, producing timestamps aligned with IceCube. Data is digitized into four channels to increase the dynamic range and read out across multiple time windows. The Flash Analog to Digital Converter (FADC) channel samples at 40\,MHz over 6.375\,$\upmu$s while the three 10-bit Analog Transient Waveform Digitizer (ATWD) channels each record 128 samples at a programmable rate (100\,--\,500\,MHz) over a \aprox 600\,ns readout window that varies among the PMTs. The data presented here is from the ATWD0, ATWD1 and ATWD2 channels, which have $\times$16, $\times$2 and $\times$0.25 gain, respectively. The ATWD channels combine for a 14-bit effective dynamic range, and events ranging from single photoelectrons to muons are recorded without saturating the data acquisition (DAQ). Single PMT threshold crossing rates are collected regularly as part of the monitoring data stream, which is recorded in addition to LC data collection. The trigger threshold for both LC and monitoring data is set to $\sim$0.25\,photoelectrons. Details of the DM-Ice17 hardware, deployment, and first data can be found in~\cite{First_data}. 

All figures shown are from Det-1. Det-2 exhibits consistent behavior and is included where it provides additional information. 

% !TEX root = Muon_PRD.tex

\section{Muon Background}
\label{sec:muons}
Atmospheric muons are primarily produced via pion decay in the upper atmosphere. The muon production rate is dependent upon the likelihood of a pion to decay relative to an interaction with atmospheric nuclei. The mean free path of atmospheric pions depends on the atmospheric density, and changes in the density with temperature cause an annual modulation in the muon production rate with a maximum in the summer \cite{PaoloGaisser, Paolo}. Muons must have an energy of several hundred GeV in order to penetrate 2200\,m.w.e.\ of ice. Muon interactions are identified in DM-Ice17  by their characteristic pulse shape and large energy depositions (\S \ref{Muon Event Identification}). The observed muon rate is 2.93\,$\pm$\,0.04\,$\upmu$/crystal/day with a 12.3\,$\pm$\,1.7\% annual modulation (\S \ref{Muon Rate and Modulation}). Muons interact as minimum-ionizing particles (MIPs) in the crystals (\S \ref{Energy Deposition}). Muon identification is verified through coincidence with IceCube (\S \ref{IceCube}), which is the subject of a separate publication in preparation. 

%%%%%%%%%%%%%%%%%%%%%%%%%%%%%%
\subsection{Muon Event Identification} 
\label{Muon Event Identification}
\begin{figure}[!htb]
	\includegraphics[width=\hw]{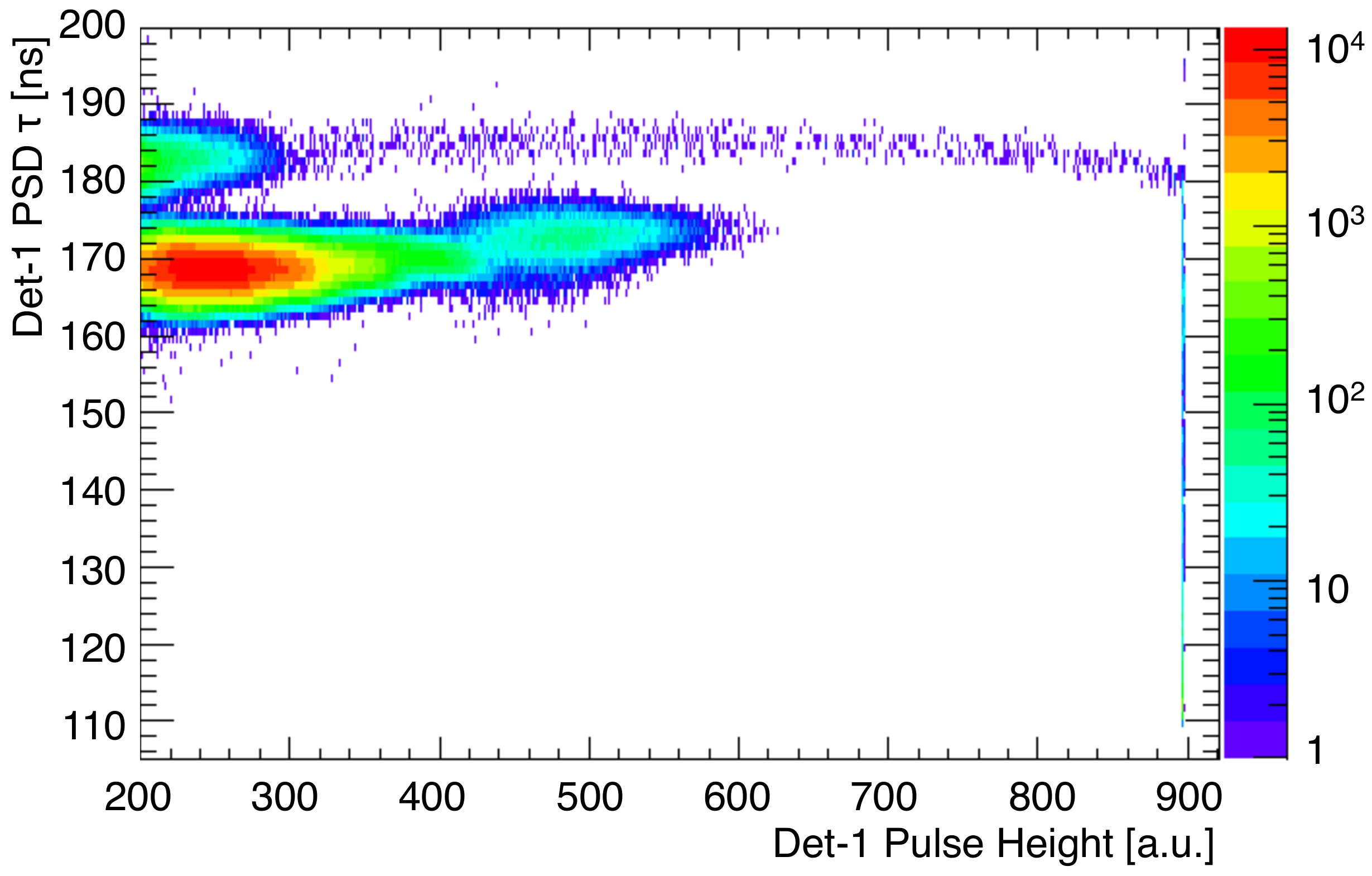}		
	\caption{Event classification in Det-1 from waveform heights and pulse shapes. Gamma events exhibit pulse heights~$<$\,325~(a.u.) and PSD $\tau$ parameters $>$\,177\,ns; alpha events comprise the lower PSD $\tau$ band, reaching up to 650~(a.u.) in pulse height. Muons make up the remaining events, with pulse heights $>$\,325~(a.u.) and PSD $\tau$ $>$\,177\,ns. The high energy pileup is due to DAQ channel saturation, with broad PSD values arising from PMT saturation. The high energy alpha population is comprised of pulses with multiple events in a single readout (e.g., bismuth-polonium) that survive PSD removal.}
	\label{Muon_tag} 
\end{figure}

\begin{figure}[!htb]
		\includegraphics[width=\hw]{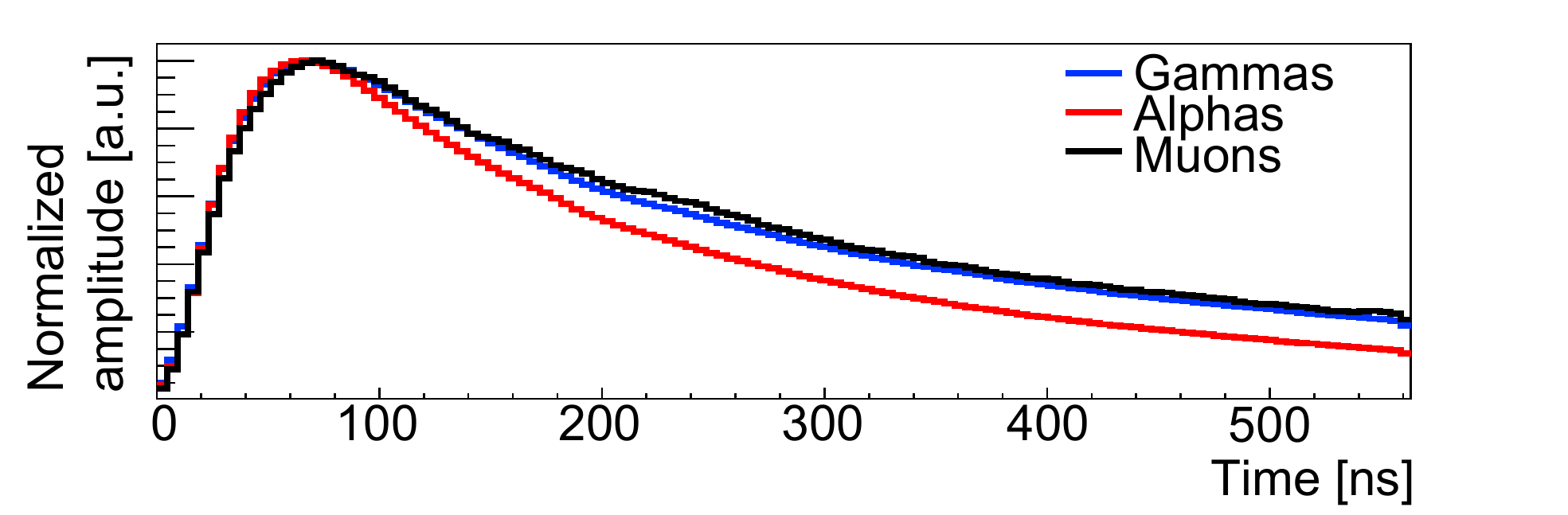}
		\caption{Average waveforms, normalized in height, of muon (black), alpha (red), and gamma (blue) events in Det-1. Alphas decay faster than gammas and muons, allowing their removal from the muon event sample. Lower energy muon events were chosen for this figure to avoid distortion from DAQ or PMT saturation.}
	\label{Average_waveform}
\end{figure}

Muon events in DM-Ice17 are separated from alpha and gamma events by a combination of pulse shape discrimination (PSD) and energy, as shown in Figure~\ref{Muon_tag}. Alpha interactions in NaI(Tl) produce pulses with a faster decay time than muon and gamma interactions \cite{DAMA_Apparatus, KIMSNaI, ANAIS_Phos2}, and they are identified through PSD, as shown in Figure~\ref{Average_waveform}. The PSD parameter, $\tau$, represents a charge-weighted mean time of the waveform. It is defined as:  
\begin{equation}
\tau = \frac{\displaystyle \sum h_i t_i}{\displaystyle \sum h_i}
\end{equation}
where $h_i$ is the height of the waveform at time $t_i$ (described in detail in~\cite{First_data}). The $\tau$ distributions are well separated, with less than one muon and one alpha event per year expected to be misidentified.

The largest energy depositions in the detector are from muon events (\S~\ref{Energy Deposition}), and isolating events above the gamma spectral cutoff at 325\,a.u. produces a highly pure muon event sample. This spectral cutoff is verified by GEANT4 simulation of the gamma backgrounds \cite{First_data}, and fitting the spectrum determines that fewer than 3 gamma events are expected in the muon sample every year. Because the muon energy deposition is proportional to the path length through the crystal, muons that clip the edge of the detector deposit less energy and are indistinguishable from gamma events. Muon path lengths through the crystal are simulated using the angular distribution of incident muons in order to determine the expected muon energy deposition. The shape of this expected distribution indicates that $<$10\% of muons appear below the gamma spectral cutoff and are misidentified as gammas. Reported muon rates correspond to those observed and do not attempt to correct for those muons that are misidentified as gammas. 

The ATWD1 channel is used to identify muons because of its higher gain. Muon energy analysis requires the ATWD2 channel due to ATWD1 channel saturation above 6.8\mevee\ (see \S \ref{Energy Deposition}). Muon selection criteria use data from a single PMT for each detector because variations in PMT response (e.g., gain, linearity, saturation energy) make single PMT analysis more effective in providing separation.

%%%%%%%%%%%%%%%%%%%%%%%%%
\subsection{Muon Rate and Modulation}
\label{Muon Rate and Modulation}
The average muon flux is consistent between Det-1 and Det-2 and is measured to be 2.93\,$\pm$\,0.04\,$\upmu$/crystal/day. The 2200\,m.w.e.\ ice overburden lowers the muon flux at DM-Ice17 by five orders of magnitude relative to the surface \cite{Bai}. Muons lose roughly 400\,GeV as they pass through the ice, and the muon flux above 1\,GeV is simulated to be 1.2~$\times$ 10$^{-3}\,\upmu$/m$^2$/s at this depth \cite{Vitaly}. This simulated flux predicts the 14\,cm diameter, 15\,cm tall crystals to detect $\sim$3 $\upmu$/crystal/day, consistent with the observed rate. 

\begin{figure}[!htb]
	\includegraphics[width=\hw]{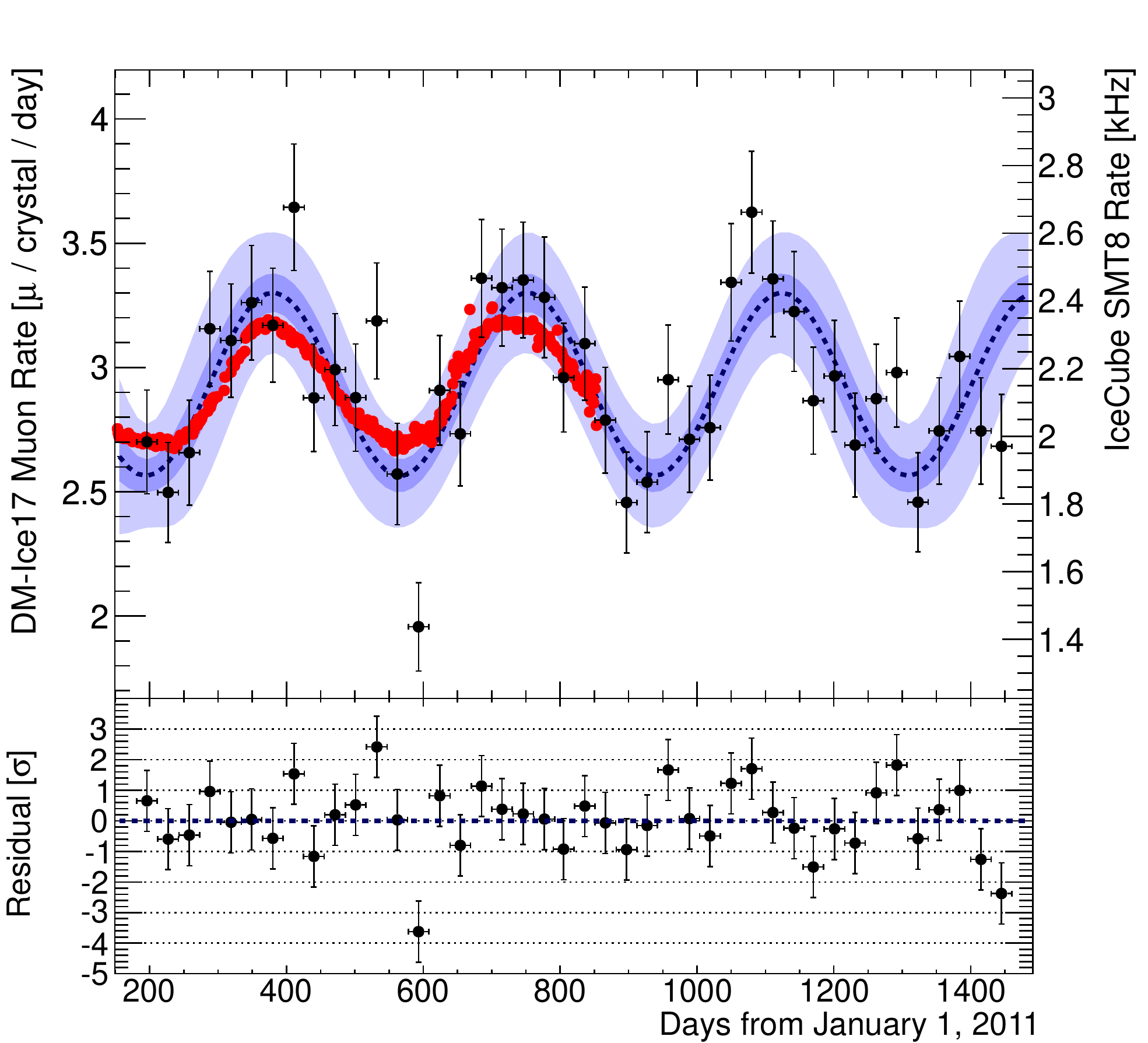}
	\caption{Combined muon rate of Det-1 and Det-2 (black circles, top figure, left axis), fit to a fixed-period sinusoid (best fit blue dashed line with 1$\sigma$ and 3$\sigma$ bands), and overlaid with IceCube rates passing an eight-fold simple majority trigger (SMT8)~\cite{ICRC2015} (red circles, right axis). DM-Ice17 data is binned by month, while IceCube data is a daily average. Statistical error on the IceCube data shown is within the size of the data point. The modulation is consistent across both DM-Ice17 detectors and fits a 12.3\,$\pm$\,1.7\% modulation amplitude. It is consistent with both the sinusoidal fit and the IceCube data, and the residual of the fit to data is shown in the lower plot. }
	\label{Initial_muons}
\end{figure}

The muon flux through DM-Ice17, $I_\upmu$, is observed to modulate with a 12.3\,$\pm$\,1.7\% modulation amplitude, as shown in Figure~\ref{Initial_muons} and Table \ref{Muon Modulation Measurements}. The rate is parameterized as an average muon flux, $I_\upmu^0$, with a modulating component of amplitude $\Delta I_\upmu$, period $T$, and phase $t_0$: 
\begin{equation}
I_\upmu = I_\upmu^0 + \Delta I_\upmu \cos \left( \frac{2 \pi}{T} (t - t_0) \right)
\label{ModEq}
\end{equation}
The modulation is expected to have a one year period and a maximum rate in the summer (i.e.,\ January at the South Pole), although small deviations from a sinusoid are expected due to the imperfectly sinusoidal form of the temperature modulation and temperature variations from year to year. Fixing a one-year period yields a modulation amplitude of $\Delta I_\upmu$~=~0.31\,$\pm$\,0.05\,$\upmu$/crystal/day with a phase of $t_0$~=~January 22\,$\pm$\,9\,days. When allowing the period to float, the best fit period is $T$~=~371\,$\pm$\,9\,days.

The modulation phase and period are consistent with those observed by IceCube~\cite{ICRC2015}, as shown in Figure~\ref{Initial_muons}. A quick rise to the maximum rate and slower decline are visible in the IceCube data; DM-Ice17 does not have enough statistics to observe this deviation from a sinusoid. The modulation fits to a fixed period sinusoid ($\chi^2$/d.o.f.\ = 50.1/39) and a floating period sinusoid ($\chi^2$/d.o.f.\ = 49.6/38) are statistically indistinguishable ($<$\,1$\sigma$ separation). A scaled fit to the IceCube data with a floating relative modulation amplitude ($\chi^2$/d.o.f.\ = 31.3/20) is preferred over a scaled fit with a fixed amplitude ($\chi^2$/d.o.f.\ = 36.2/21) to greater than 2$\sigma$. All are preferred to greater than 5$\sigma$ over the null hypothesis. 

\begin{table*}[!htb]
\caption{The average rate ($I_\upmu^0$), modulation amplitude ($\Delta I_\upmu$), phase ($t_0$), and best fit period ($T$) (see Eq. \ref{ModEq}) of the muons in DM-Ice17. The modulation amplitude is reported in [$\upmu$/crystal/day] and in [\%]. The amplitude and phase are from a fixed one-year period modulation, with the best fit period reported in the right column. These values are consistent between detectors and agree with expectation.}
\label{Muon Modulation Measurements}
\begin{ruledtabular}
\begin{tabular}{p{2.4cm}|ccccc|c}
            & \multicolumn{5}{c|}{\bf{Fixed Period Fit}}									& \bf{Floating Period Fit} \\
            & \bf{$I_\upmu^0$}	& \multicolumn{2}{c}{\bf{$\Delta I_\upmu$}}	& \bf{$t_0$}	&		& $T$ \\
            & [$\upmu$/crystal/day]	& [$\upmu$/crystal/day]		& [\%]		& [days]		&		& [days] \\
\hline
\bf{Det-1}		& 2.89\,$\pm$\,0.05		& 0.39\,$\pm$\,0.07		& 13.5\,$\pm$\,2.4\%		& Jan 22\,$\pm$\,11	&& 387\,$\pm$\,16 \\
\bf{Det-2}		& 2.96\,$\pm$\,0.07		& 0.33\,$\pm$\,0.07		& 11.1\,$\pm$\,2.4\%		& Jan 21\,$\pm$\,12	&& 370\,$\pm$\,14 \\
\bf{Combined}	& 2.93\,$\pm$\,0.04		& 0.36\,$\pm$\,0.05		& 12.3\,$\pm$\,1.7\%		& Jan 22\,$\pm$\,9	&& 371\,$\pm$\,9 \\
\end{tabular}
\end{ruledtabular}
\end{table*}

The amplitude of the annual modulation in the atmospheric temperature and the corresponding muon rate are maximal at the South Pole due to its geographic location; daily temperature fluctuations do not mitigate the annual variation. The muon production mechanism induces a modulation amplitude that increases with energy, leading to a depth dependent amplitude; higher energy muons increasingly match the fractional modulation of the temperature modulation, which is 11\% at the South Pole~\cite{PaoloGaisser}. The muon modulation has been measured by IceCube (1450\,--\,2450\,m) at the South Pole to be 8.6\,$\pm$\,1.2\% \cite{Paolo}. DM-Ice17's location at the bottom of IceCube means that only higher energy muons reach it, leading to a larger modulation amplitude that corresponds more closely to that of the atmospheric temperature modulation~\cite{PaoloGaisser}. 

%%%%%%%%%%%%%%%%%%%%%%%%%
\subsection{Energy Deposition}
\label{Energy Deposition}
Simulations indicate that the muons reaching DM-Ice17 are most likely to exhibit a MIP energy loss as they pass through the crystal (\aprox7\,MeV/cm) with total energy deposition proportional to track length. Accounting for the predicted muon angular distribution and the crystal geometry, the muon path lengths range from 0 to 21\,cm with a strong peak at 15\,cm. This corresponds to an expected muon energy distribution peaked at 100\,MeV with a flat continuum extending down to 0\,MeV.

\begin{figure}[!htb]
	\includegraphics[width=\hw]{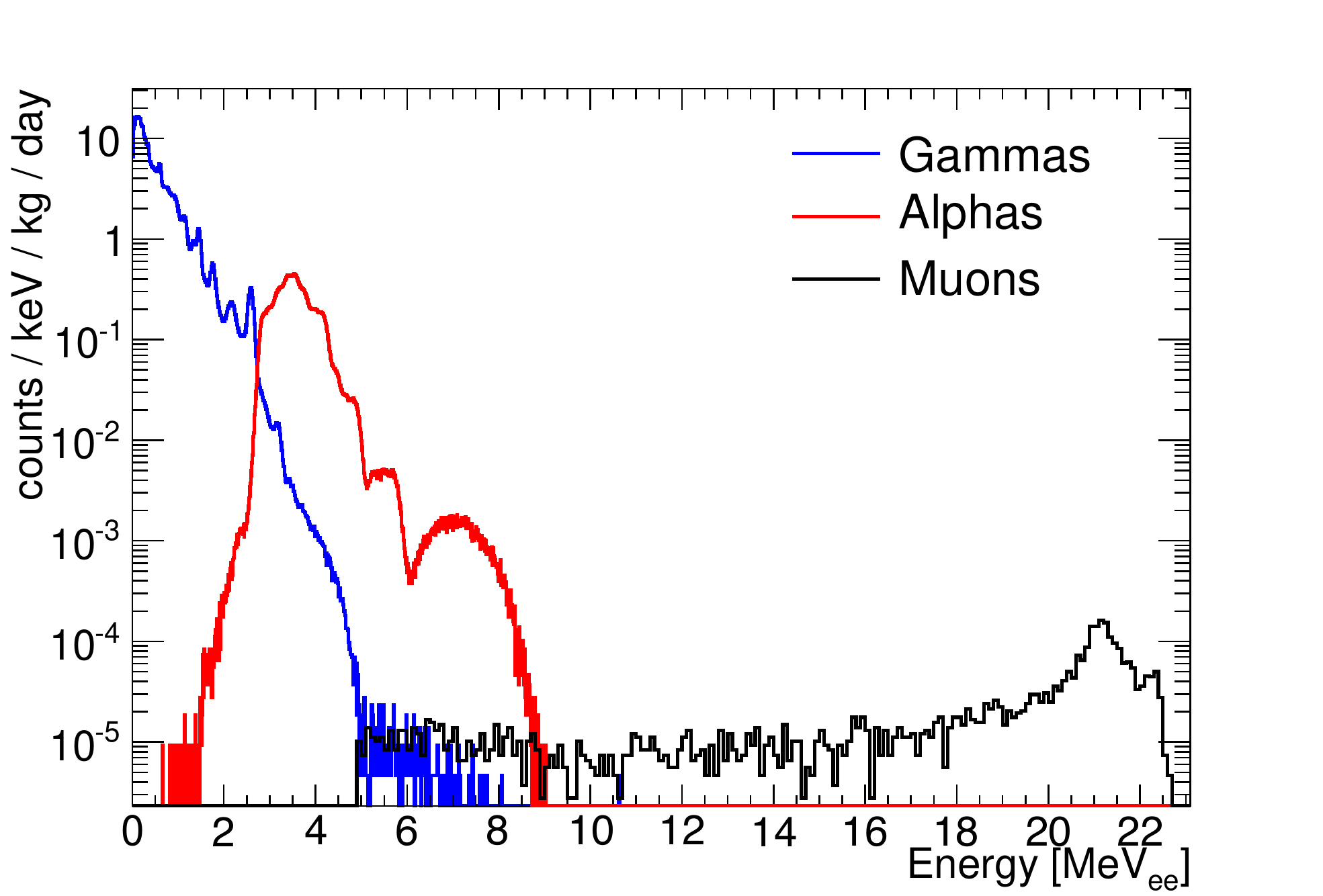}  
	\caption{The energy spectrum of Det-1 gamma (blue), alpha (red), and muon events (black). Muons produce the highest energy events in the detector, extending out to 23\mevee. Detector response effects produce a reconstructed energy that is significantly smaller than the simulated energy deposition of $\sim$100\,MeV. The highest energy alpha peak ($\simeq$\,7.5\mevee) is from bismuth-polonium decays.}
	\label{Mu spectrum}
\end{figure}

The energy spectrum of Det-1 is shown in Figure~\ref{Mu spectrum} with events ascribed to gammas, alphas, and muons as detailed in \S \ref{Muon Event Identification}.  A linear energy calibration is established from the locations of gamma peaks out to 2.6\mevee~\cite{First_data}.  Integrated charge in the ATWD1 channel is used for the energy scale of alphas and gammas.  At higher energies, the integrated charge no longer exhibits a monotonic relationship with energy due to the effects of DAQ and PMT saturation; therefore, pulse height in the ATWD2 channel is used for muons. Linear extrapolation of the gamma peak height calibration results in an observed muon spectrum extending to only 23\mevee, despite the expectation of a spectrum peaked at 100\,MeV. Uncorrected PMT saturation effects are responsible for the difference.

%%%%%%%%%%%%%%%%%%%%%%%%%
\subsection{IceCube Coincidence}
\label{IceCube}
DM-Ice17 muon identification is verified through observed coincidence with IceCube. Det-1 is located in the center of the IceCube area, with full azimuthal acceptance; Det-2 is located on the outer edge of IceCube, limiting its solid angle acceptance. Of muon events identified in Det-1 over the 2012\,--\,2013 run year, 93\% were coincident with muon events in IceCube; 43\% of muon events in Det-2 were coincident over this same period. These coincidence rates are consistent with expectation once IceCube data selection efficiency is taken into account. This agreement with expectation verifies the low leakage of gamma and alpha events into the muon sample. 

% !TEX root = Muon_PRD.tex
\section{Phosphorescence} 
\label{sec:phosphorescence}
NaI(Tl) crystals emit photons primarily in fluorescent scintillation but also release photons from longer excitations, known collectively as phosphorescence or afterglow (\S \ref{Phosphorescence in NaI(Tl)}). Phosphorescence is broadly classified as shorter-lived ($\lesssim$\,ms) and longer-lived ($>$\,ms) excitation. Muon events in DM-Ice17 induce long-lived phosphorescence, increasing in magnitude with muon energy. Comprised of up to tens millions of photons, these phosphorescent periods last for tens of seconds (\S \ref{Muon-Induced Phosphorescence}). Deadtime in DM-Ice17 precludes a statement on short-lived phosphorescence.

%%%%%%%%%%%%%%%%%%%%%%%%%%%%%%%%%%%%%%%%%%%%%%%%%
\subsection{Phosphorescence in NaI(Tl)}
\label{Phosphorescence in NaI(Tl)}
Phosphorescence refers to NaI(Tl) scintillation longer than the $\sim$250\,ns primary fluorescent decay component~\cite{Knoll}. A suite of NaI(Tl) experiments has observed phosphorescence with timescales ranging from 1.5\,$\upmu$s to 45\,days~\cite{Phos7, ANAIS_Phos, ANAIS_Phos2, KIMSNaI, StGobain, Phos6, SecsPhos, French, Fronka, Phos8, Reactor, DaysPhos, Cuesta_thesis}. Differences in experimental results indicate numerous phosphorescence decay channels. No correlations between phosphorescence decay time and crystal size or temperature have been observed in these studies. Phosphorescence has been observed following exposure to cosmic rays, radioactive sources, and ultraviolet (UV) and visible light. It is determined that the initial intensity of decays shorter than the second-scale is proportional to the intensity of incident radiation, while that of longer decays (minutes to hours) is proportional to the absorbed dose~\cite{Phos7}. Both the ANAIS and KIMS dark matter experiments have measured elevated trigger rates following large energy depositions. ANAIS reports a 70\,--\,100\,ms~\cite{ANAIS_Phos, ANAIS_Phos2} and a minutes-long \cite{Cuesta_thesis} decay, and KIMS reports a few seconds decay~\cite{KIMSNaI}. St.\ Gobain, the manufacturer which produced the DAMA crystals, has measured a phosphorescent component over hours to days following exposure to UV light~\cite{StGobain}. By contrast, DAMA has not reported any observation of phosphorescence in their analysis data sample. No details of a phosphorescence study have been disclosed, but it has been reported that no events in the dark matter signal region are attributed to phosphorescence. Phosphorescence is generally comprised of low energy events, with the St.\ Gobain reporting phosphorescence reaching 10\,keV~\cite{StGobain}. For a detailed discussion of experimental results, see~\cite{Thesis}. 

The precise mechanisms of longer ($>$\,ms) phosphorescent decays are not entirely understood, but they are postulated to include long-lived metastable activator states and traps from crystal defects and impurities, both intrinsic and irradiation-induced. In the metastable interpretation, an electron excites the activator center to a state that is forbidden to decay to ground and requires additional energy, generally from thermal excitation, to reach a state that can decay. The decay time is then characteristic of the metastable state lifetime. In the trap interpretation, the charge carriers are trapped and thermally released to recombine; this interpretation can predict trap depths and equilibrium conditions, as described in detail in~\cite{Phos8}. Shorter phosphorescent decays are understood as emerging from self-trapped exciton emission, lasting $\upmu$s\,--\,ms, and a long-lived recombination channel that produces $\upmu$s-long decays~\cite{Rodnyi}. Long-lived recombination occurs when an electron and hole reach different thallium (Tl) activator sites, and the electron must be thermally released before recombination~\cite{Diffusion, Rodnyi}. The phosphorescence intensity, spectral composition, and decay time are dependent upon the purity of the raw material, growing conditions, and irradiation exposure~\cite{Rodnyi}. Phosphorescence is highly variable across crystals and must be studied individually for each detector. 

%%%%%%%%%%%%%%%%%%%%%%%%%%%%%%%%%%%%%%%%%%%
\subsection{Muon-Induced Phosphorescence in DM-Ice17}
\label{Muon-Induced Phosphorescence}
Muon interactions can induce phosphorescence in the DM-Ice17 crystals. Phosphorescence manifests as an increase in the number of uncorrelated single photons emitted directly following a muon event in DM-Ice17. Increases of up to tens of millions of photons have been detected. Phosphorescence is observed in both LC data, which requires coincidence between the two PMT coupled to the same crystal within 350\,ns, and single PMT threshold crossings registered in monitoring data (see \S \ref{sec:dmice}).   

The monitoring data stream, rather than the LC data stream, is used to determine phosphorescence decay parameters because it is a more direct measurement of photon emission rates. This analysis was performed on an eight month subset of the data (July 1, 2011\,--\,February 28, 2012), during which monitoring data was recorded every 2\,s. Subsequent data was monitored every 60\,s \cite{First_data}, limiting its ability to study phosphorescence. The photon rate in the monitoring stream is determined by counting the number of threshold crossings over a 900\,ms period. Photon rates for this study are determined by averaging the first two monitoring data records in each PMT after a muon event, which corresponds to averaging across the 4\,s following a muon interaction. 

\begin{figure}[!ht]
	\includegraphics[width=\hw]{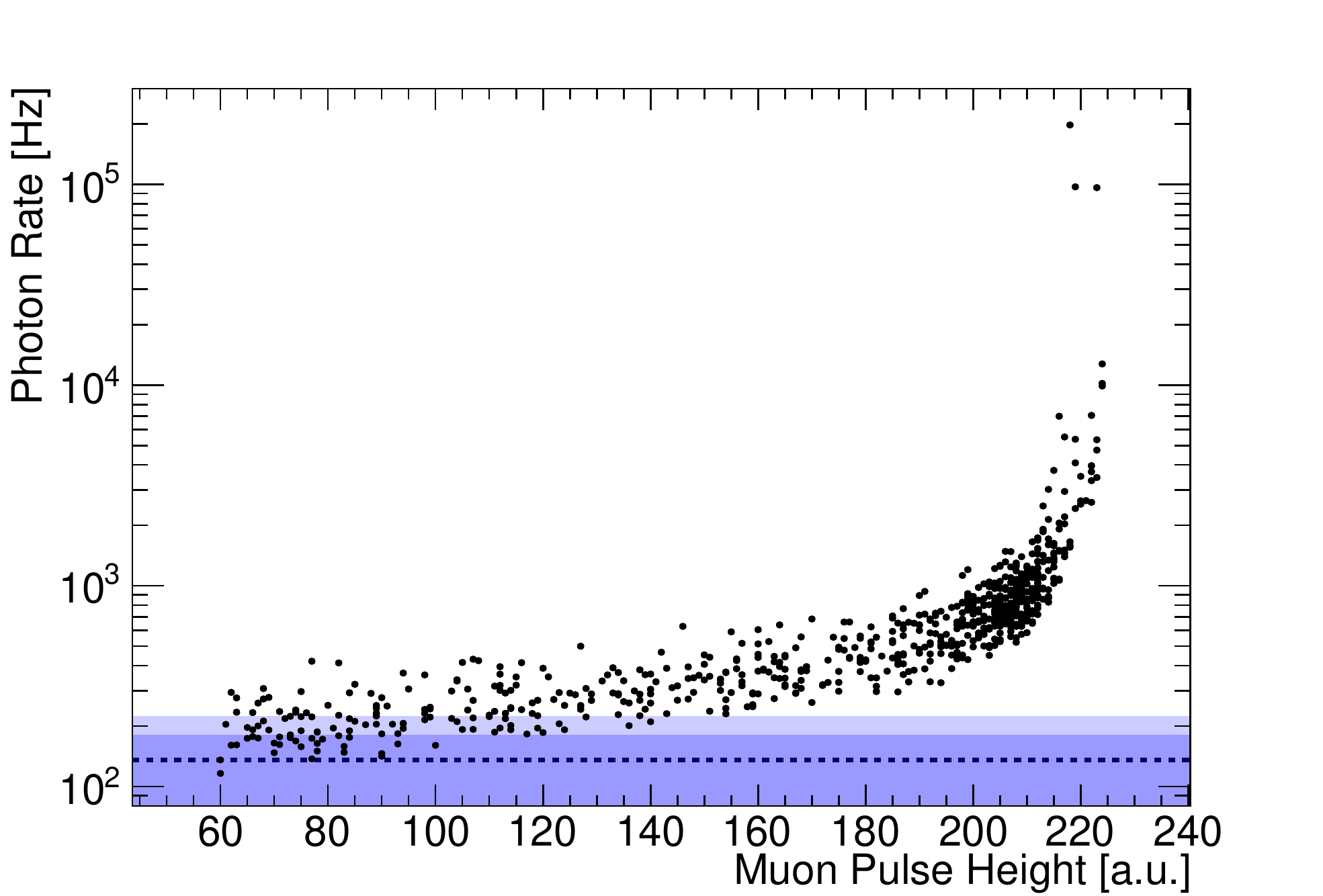}
	\caption{The average photon rate from the first two monitoring data records following a muon event, averaged across both PMTs in Det-1 (black) and compared to the average photon rate (blue dashed), shown with single-sided bands encompassing 90\% (dark blue) and 99\% (light blue) of data. The energy deposition is determined by the pulse height of the muon event in a.u. Data is from the eight month period when monitoring data was taken at 2\,s intervals. Det-2 shows consistent behavior.} 
	\label{Phos_cause} 
\end{figure}

\begin{figure}[!ht]
	\includegraphics[width=\hw]{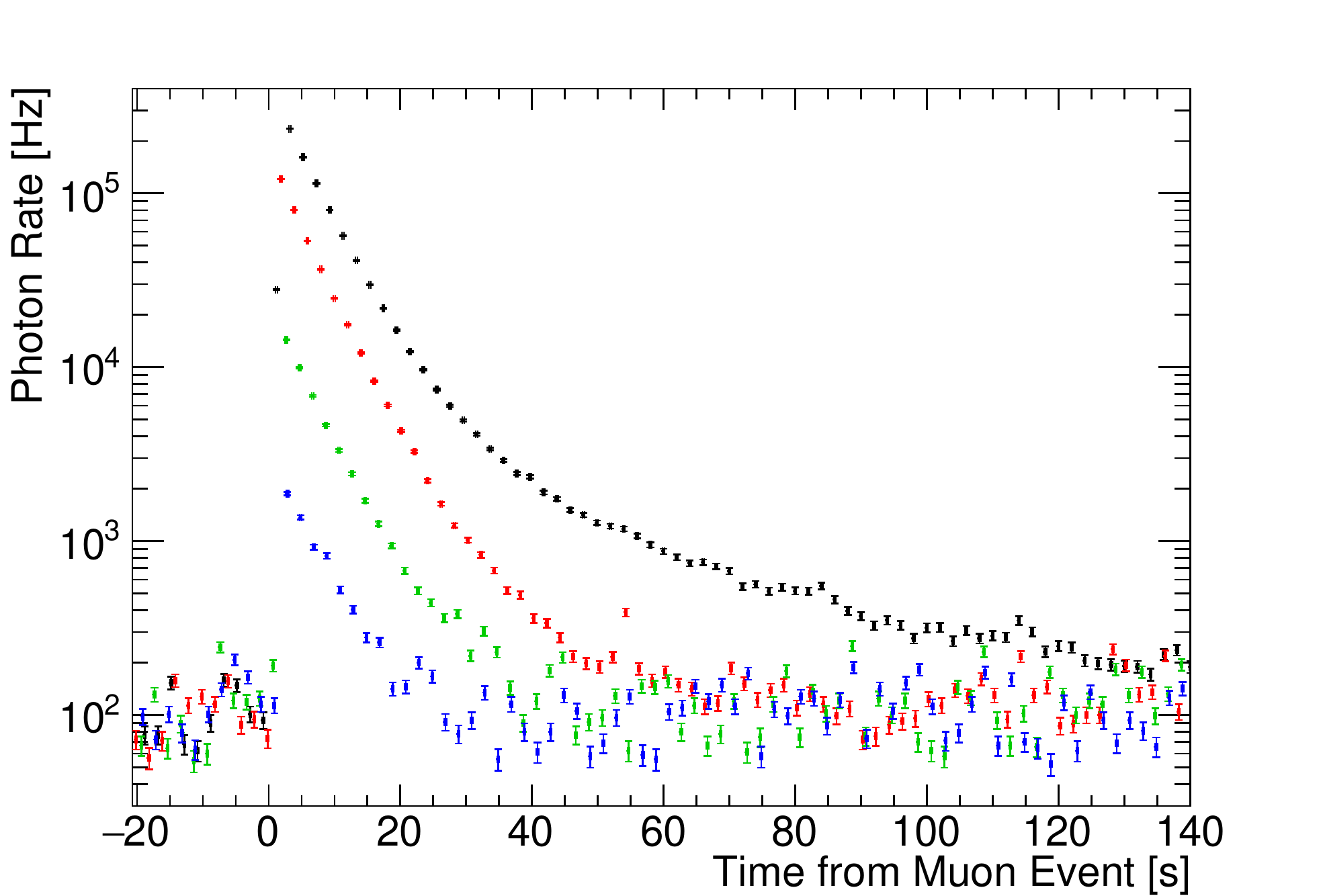}
	\caption{Muon-induced phosphorescence is observed as elevated photon rates that decay with a characteristic decay time of 5.5\,$\pm$\,0.5\,s. Each color corresponds to a separate phosphorescence period in Det-1 monitoring data. The x-axis is set to zero at the time of the muon event inducing phosphorescence. The largest phosphorescence periods, including the black and red curves shown, exhibit a second decay time of $\sim$30\,s. Det-2 shows consistent behavior. } 
	\label{PhosComps} 
\end{figure}

A statistically significant event rate increase is induced by 96\% of muons, which induce rates higher than observed in 90\% of normal running distributions. The magnitude of phosphorescence increases with muon energy, as shown in Figure \ref{Phos_cause}. Trigger rates during phosphorescence can spike outside of the fluctuations of average trigger rates, which are 2.5\,Hz/crystal in coincident data and 100\,Hz/PMT in monitoring data. Single PMT threshold crossings in the monitoring stream have been observed to reach as high as 230\,kHz during phosphorescence. The photon rate increases immediately following the muon events and decays exponentially with a 5.5\,$\pm$\,0.5\,s decay time, as shown in Figure \ref{PhosComps}. The brightest phosphorescence events ($<$1\% of phosphorescence) exhibit a second resolvable decay time of $\sim$30\,s. 

The high rate of phosphorescent photon emission leads to accidental LC triggers during large phosphorescence periods. Phosphorescence induces statistically significant LC rate increases following 2\% of muons, when they are identified in LC data by recording the number of triggers in the 30\,s after the muon and comparing this to a baseline rate defined as the number of triggers in the 30\,s before the muon. The elevated LC trigger rates during phosphorescence exceed the designed range for the DAQ. The 730\,$\upmu$s deadtime that follows each event \cite{Walter} and the scale of observed phosphorescence prevents the detection of the ms-scale decays reported by other experiments \cite{ANAIS_Phos, ANAIS_Phos2}. An additional deadtime is introduced above a LC rate of 50\,Hz, which only occurs during phosphorescence. Analysis of these considerations and inclusion of quantum efficiency indicates that only a fraction of photons are recorded in most phosphorescence events, falling with increasing trigger rate to only 0.03\% of photons recorded in the largest, DAQ saturated events. Of the $\sim$10$^8$ photons emitted in the largest phosphorescent event, 3940 triggers were recorded. When corrections for these effects are implemented, the photon rate extrapolated from LC data is consistent with the recorded monitoring rate. LC phosphorescence data from the entire 3.5\,yr dataset is used for analysis, and it is consistent before and after the monitoring frequency change. 

\begin{figure}[!ht]
	\includegraphics[width=\hw]{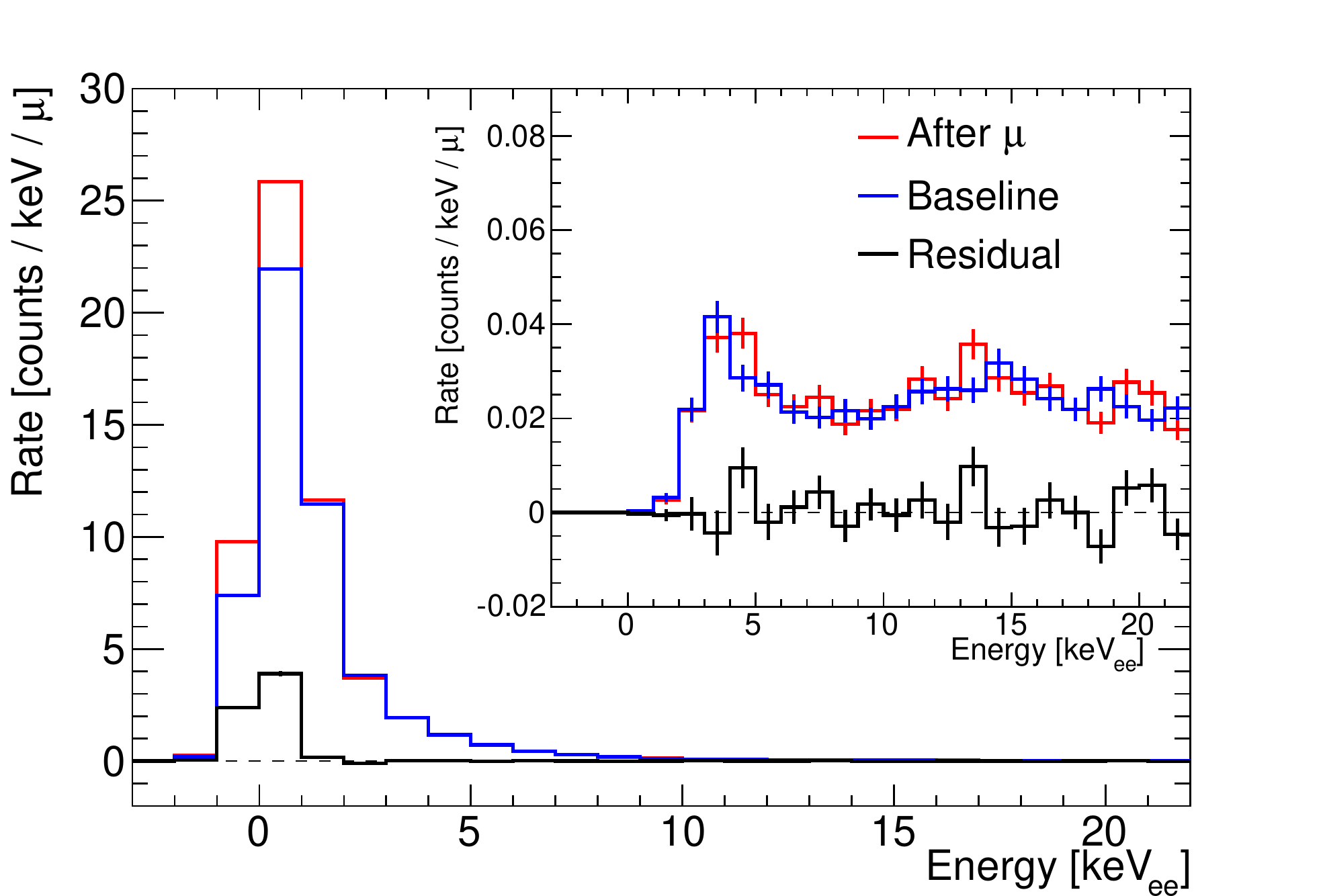}
	\caption{ATWD0 energy spectrum of phosphorescent LC triggers in Det-1 (black), isolated by subtracting the baseline spectrum derived from all pulses in the 30\,s before the muon (blue) from the spectrum of the 30\,s after the muon (red). Muons that saturate the DAQ or are within 30\,s of a run transition have not been included. The residual indicates the additional events due to phosphorescence, which appear as low energy noise. The inset shows the spectral comparison, with statistical error bars, after the application of noise cuts and illustrates the rejection of phosphorescence events with noise removal.} 
	\label{PhosEnergy} 
\end{figure}

Phosphorescence events are primarily comprised of discrete, low energy pulses that are dominated by single photoelectron-like triggers and are rejected as noise. Det-1 (Det-2) has a light yield of 5.9 (4.3) photoelectrons/keV, and energy is defined as the integrated charge over the $\sim$600\,ns readout window. Figure~\ref{PhosEnergy} shows the energy spectrum of all LC phosphorescence triggers in Det-1, isolated by subtracting a baseline spectrum from the spectrum of all pulses in the 30\,s after the muon. The baseline spectrum is made from all pulses in the 30\,s before the muon. LC phosphorescence triggers produce a spectrum that is peaked below 1\kevee\ with no statistically significant component above 2\kevee\, even before noise rejection. Phosphorescence events do not mimic normal scintillation events and are therefore rejected by the standard noise removal procedures described in~\cite{First_data}. This rejection leaves a spectrum that is consistent with the average, non-phosphorescence background spectrum at all energies. Muon-induced phosphorescence therefore does not impact the dark matter analysis sample in DM-Ice17.

Phosphorescence is consistent across both PMTs on each crystal and across PMT high voltage settings in DM-Ice17. This indicates that it is not the result of an effect from the PMT dynode chain. 

% !TEX root = Muon_PRD.tex
\section{Implications for Sodium Iodide Experiments}
\label{Implications}

Phosphorescence induced by muons has been proposed as a potential modulating background in the region of interest for NaI(Tl) dark matter experiments~\cite{Nygren, Davis, Blum}. Such an effect requires a sufficient photon rate following a muon to produce photoelectron pileup that reconstructs above 2\kevee.  Muon-induced phosphorescence is not rejected by traditional muon tagging and can evade multi-crystal and energy cuts. This must be considered when assessing the feasibility of muons and muon-induced phosphorescence to mimic a low energy scintillation-like signal. 

The observed phosphorescence in DM-Ice17 indicates that this is unlikely to be the sole source of the annual modulation signal observed in DAMA/NaI and DAMA/LIBRA, consistent with those experiments' statements \cite{No_role, No_role2, DAMAProceedings}, although variations in detector details and in phosphorescence across crystals complicate a comparison between experiments. In order for phosphorescence to be the sole source of DAMA signals, the events in the region of interest must be entirely muon-induced, and this is highly improbable~\cite{No_role, DAMAProceedings}. Applying the phosphorescence observed in DM-Ice17 to DAMA's muon background indicates that the event rate would be too small to produce the observed modulation, even before noise removal. Production of the DAMA modulation signal would require a significantly higher number of phosphorescent photons with a time structure conducive to photoelectron pileup. Additionally, the phase of the modulation would have to precede that of muons by 45\,days~\cite{Borexino, DAMAFinal}. A second annually modulating background with a different phase could alter the apparent amplitude and phase of the combined signal \cite{Davis}, but no such background has been established~\cite{No_role2, Klinger}.

DAMA has reported that muons and associated phosphorescence cannot produce their observed annual modulation signal~\cite{No_role, No_role2, DAMAProceedings}. The observations of DM-Ice17 presented here provide no indication that muon-induced phosphorescence is responsible for the modulating signal observed in DAMA/NaI and DAMA/LIBRA.  However, variations in the properties of phosphorescence among crystals~\cite{ANAIS_Phos, Rodnyi} require that each NaI(Tl) experiment study the possible signal leakage of its phosphorescence background.

% !TEX root = Muon_PRD.tex

\section{Conclusions}
\label{sec:conclusions}
We have identified and characterized muons and muon-induced phosphorescence in DM-Ice17. Muon events are identified by their pulse shape and energy deposition, yielding an average observed flux of 2.93\,$\pm$\,0.04~$\upmu$/crystal/day.  The muon rate, correlated to atmospheric temperature, modulates with a 12.3\,$\pm$\,1.7\% amplitude and a phase corresponding to a maximum flux on January 22\,$\pm$\,9\,days. Muon identification is validated with observed coincidence in IceCube, indicating a highly pure muon sample in DM-Ice17.  Muon events induce a long-lived phosphorescence that is comprised of up to tens of millions of photons. The rate of photons from phosphorescence falls off exponentially with a 5.5\,$\pm$\,0.5\,s decay time. While this effect provides a mechanism for muons to induce a modulation at low energies that escapes muon removal cuts, the phosphorescence pulses observed in DM-Ice17 appear primarily below 2\kevee. These events are rejected as noise, and they do not have the event rate in DM-Ice17 required to produce a modulation consistent with the DAMA signal, even before noise removal. Variations in detector details and in the properties of phosphorescence between crystals render an inter-experimental comparison difficult and motivate detailed studies by all experiments on the potential impact of phosphorescence in the region of interest. We conclude, however, that the annually-modulating signal reported by the DAMA experiments cannot be accounted for by phosphorescence with the specific characteristics observed in DM-Ice17.

% !TEX root = Muon_PRD.tex

\section{Acknowledgments}
We thank the Wisconsin IceCube Particle Astrophysics Center (WIPAC) and the IceCube collaboration for their on-going experimental support and data management, Benedikt Riedel for assisting in the implementation of the IceCube rate information, and Paolo Desiati for useful conversations about muon modulations. This work was supported in part by the Alfred P.\ Sloan Foundation Fellowship, NSF Grants No.~PLR-1046816, PHY-1151795, and PHY-1457995, WIPAC, the Wisconsin Alumni Research Foundation, Yale University, the Natural Sciences and Engineering Research Council of Canada, and Fermilab, operated by Fermi Research Alliance, LLC under Contract No.~DE-AC02-07CH11359 with the United States Department of Energy. W.~P.~and A.~H.~were supported by the DOE/NNSA Stewardship Science Graduate Fellowship (Grant No.~DE-FC52-08NA28752) and NSF Graduate Research Fellowship (Grant No.~DGE-1256259) respectively.

%%%%%%%%%%%%%%%%%%%%%%%%%%%%%%%%%%%%%%%%%%%%%%%%%%%%%%%%%%%%
%%  References with bibTeX database:
%%%%%%%%%%%%%%%%%%%%%%%%%%%%%%%%%%%%%%%%%%%%%%%%%%%%%%%%%%%%
%%  natbib.sty is loaded by default. However, natbib options can be
%%  provided with \biboptions{...} command. Following options are
%%  valid:
%%%%%%%%%%%%%%%%%%%%%%%%%%%%%%%%%%%%%%%%%%%%%%%%%%%%%%%%%%%%
%%   round  -  round parentheses are used (default)
%%   square -  square brackets are used   [option]
%%   curly  -  curly braces are used      {option}
%%   angle  -  angle brackets are used    <option>
%%   semicolon  -  multiple citations separated by semi-colon
%%   colon  - same as semicolon, an earlier confusion
%%   comma  -  separated by comma
%%   numbers-  selects numerical citations
%%   super  -  numerical citations as superscripts
%%   sort   -  sorts multiple citations according to order in ref. list
%%   sort&compress   -  like sort, but also compresses numerical citations
%%   compress - compresses without sorting
%%%%%%%%%%%%%%%%%%%%%%%%%%%%%%%%%%%%%%%%%%%%%%%%%%%%%%%%%%%%

\addcontentsline{toc}{section}{References}
\medskip

\bibliographystyle{apsrev4-1}

\bibliography{%
biblio%
}

\end{document}